\documentclass[12pt, letterpaper]{article}

\usepackage{amsmath,amssymb}
\usepackage{cite}
\usepackage{fancyhdr}
\usepackage[top=1in, bottom=1in, left=1in, right=1in]{geometry}
\usepackage{graphicx}
\usepackage{hyperref}

\numberwithin{equation}{section}
\date{}

\begin{document}
\title{{\rm\footnotesize \qquad \qquad \qquad \qquad \qquad \ \qquad \qquad \qquad \ \ \ \ \ \                              RUNHETC-2026-3}\vskip.5in  What does it mean to have a quantum gravitational theory of de Sitter Space? \\  Submitted to 2026 Gravitation Research Foundation Essay Contest Jan 1, 2026}
\author{Tom Banks (corresponding author)\\
NHETC and Department of Physics \\
Rutgers University, Piscataway, NJ 08854-8019\\
E-mail: \href{mailto:tibanks@ucsc.edu}{tibanks@ucsc.edu}
\\
\\
}

\maketitle
\thispagestyle{fancy} 

\begin{abstract} We argue that if de Sitter space is indeed represented by a finite dimensional quantum system, then semi-classical considerations, combined with the fundamental principles of quantum measurement theory, imply that any theoretical model of it is ambiguous. If our own universe asymptotes to such a de Sitter state, and if a model of it as a finite system can be embedded in a sequence of models that converges to a non-perturbative completion of a unique superstring model in asymptotically flat space, then one might be able to find a very precise mathematical model of our universe. However, even the most comprehensive experiments possible to local detectors in the universe cannot measure more than a tiny fraction of the total number of q-bits in the system. \normalsize \noindent  \end{abstract}


\vspace{1cm}

\vfill\eject


\section{What do we know?}

de Sitter (dS) space is the maximally symmetric solution of Einstein's equations with positive cosmological constant.  Although it was not always so, most researchers have now come around to the hypothesis that a quantum theory of this space-time must view it as a finite entropy system, with the empty dS state having more entropy than any state with localized excitations\cite{tbwfds}\cite{tb2001}\cite{bfm}.   There is less agreement on some other old claims\cite{ckn}\cite{tbwfsp}, which suggest that one cannot {\it in principle} build a localized detector that can measure more than a fraction $S_{dS}^{(d - 1)/(d)}$ of the q-bits in the dS Hilbert space.  This means that a model of dS space with a Hilbert space of dimension $e^{S_{dS}}$ can never be fully verified by any conceivable experiment.  {\it A fortiori} any model with an infinite dimensional operator algebra is massive overkill.  

Another fundamental flaw of many attempts to build theoretical models of dS space comes from the fact that the masses of localized objects in dS space are always bounded.  As a consequence, no localized object can remain close to a geodesic in dS space for a time longer than of order $R_{dS} {\rm ln}\ \frac{R_{dS}}{L_P} . $ This means that a quantum theory of dS space based on a fundamental time independent Hamiltonian does not make any sense. 

We will argue that there are two avenues towards building models of dS space that have some scientific or mathematical validity.  The first is to concentrate on the measurements possible to the most robust, long lived detector possible in a given dimension, and perhaps with a given matter content.  The resulting theory will depend on the idiosyncrasies of the detector's trajectory and composition.  We will describe examples below.  The second possible direction is to build models with a variable cosmological constant (c.c.) parameter, which converge in the limit of zero c.c. to some well defined superstring model in asymptotically flat space.  The evidence from string theory suggests that this can only occur if the limiting model is exactly supersymmetric and the number of asymptotically flat dimensions is four.  In general, the first strategy leads to a wide range of possible models for any given classical solution, because of the {\it in principle} limitations on any model of a localized detector in dS space.  The second might allow for more precise model building, once a particular asymptotically flat $4$ dimensional string theory model is known and one invents a construction that defines its non-perturbative S-matrix.  

\section{Two Dimensional Models}

The most general form of dilaton gravity in $1 + 1$ dimensions can be written
\begin{equation} I = \int d^2 x \sqrt{-g} [  S R  - V(S) - K(S) g^{mn} \partial_m S \partial_n S] . \end{equation} 
Assume a solution with an exact dS metric and $S(t)$ depending only on the time in global coordinates.  Then the Hamiltonian constraint reads.
\begin{equation} T(t) \dot{S} + \frac{K(S)}{2} (\dot{S})^2 + \frac{1}{2} V(S) = 0, \end{equation} 
where $T(t) \equiv R_{dS}^{-1} \tanh (t / R_{dS}) . $   So $ T \rightarrow \pm R_{dS}^{-1}$ as $ t \rightarrow \pm \infty $.  The Hamiltonian constraint then implies that if $S$ becomes a constant $S_0$ at infinity and $K (S_0)$ does not blow up, we must have $V(S_0) = 0$.

On the other hand, the scalar field equation says that
\begin{equation} R + {\rm terms\ involving\ } \dot{S} = V^{\prime} (S) . \end{equation}
Since the scalar curvature is positive, this is only possible if $V^{\prime} (S) \rightarrow$ constant at large time , but this is incompatible with $V(S_0) \rightarrow 0$.  
We conclude that, in global coordinates $|S(t)|$ must go to infinity when $|t|$ goes to infinity.

Many models of dilaton gravity arise by dimensional reduction of higher dimensional models.  If one starts in Einstein conformal frame in higher dimensions, then $S$ always has the meaning of the volume of a compact manifold, and one can view it, up to an additive constant, as an entropy.  This suggests that none of these models can be thought of as having to do with a finite entropy dS space, but one must be careful.

Among these models we find the dimensional reduction of pure higher dimensional gravity with positive c.c..  In higher dimensions, the infinite radius of a global two sphere has nothing to do with infinite entropy.  If we consider the causal diamond along any time-like trajectory, it has finite area in the limit of infinite proper time.  In $1 + 1$ dimensions we have no independent measure of the entropy of a diamond in the spatially homogeneous dS geometry.  In\cite{jtdS} we interpreted most of the classical solutions of the Jackiw-Teitelboim dS gravity model as describing the Coleman Deluccia decay of an unstable excitation of the static patch solution into an infinite entropy Big Bang cosmology.  We constructed an infinite number of different quantum mechanical models whose hydrodynamics agreed with the gravity description.  A similar construction is possible for all dS solutions of $1 + 1$ dimensional dilaton gravity.  Depending on the properties of $K(S)$ the roles of the origin and the cosmological horizon in the decay process can be reversed.  There is also one solution of JTdS gravity where the entropy at one pole of the Euclidean sphere is zero.  This solution is the dimensional reduction of $2 +1$ dimensional dS space and does not have a tunneling interpretation.

Our point here is not that the interpretation proposed in\cite{jtdS} is correct, but rather that, absent other information the classical gravitational action and its solutions do not provide sufficient data to pin down a precise quantum mechanical model.  It's quite clear that the dimensional reduction of $3$ or $4$ dimensional dS space could refer simply to a subsector of the models we'll describe below, but given only its $1 + 1$ dimensional action, there are an infinite number of other possible quantum models that it could describe.  

\section{Three Dimensions}

Three dimensional dS space, with spherical spatial sections in global coordinates is a particularly simple space-time because of two special properties.  Like dS space in higher dimensions, localized excitations decrease the area of the horizon and there's a maximal total mass. Unlike higher dimensions, there are no gravitational bound states.  This means that any model of massive quantum field theory coupled to gravity is making serious errors from the outset.  We cannot make mathematical models of complex detectors\footnote{The problem of studying the quantum electrodynamics of finite numbers of charged relativistic massive particles, coupled to gravity has not been solved.   It's possible that one could construct complex detectors in such a model.  However, it is not even clear that the coupled system of pure electromagnetism and gravity is a consistent quantum theory.}.  If one studies a first quantized model of relativistic particles in $dS_3$, one finds\cite{satbwf} that a localized wave function spreads over the horizon of any static patch in a time no longer than $R_{dS} {\rm ln} \frac{R_{dS}}{L_P}$, so there is no justification for using the static patch time to define time evolution.  Since the particles themselves have no internal structure which could decohere a ``clock", there is no real principle with which to choose a Hamiltonian.  In\cite{satbwf} we constructed a family of time dependent Hamiltonians on a finite dimensional Hilbert space, which fit much of the semi-classical data of relativistic particle quantum mechanics in $dS_3$.  It seems doubtful that one could do anything more, unless one finds a way to connect a sequence of $dS_3$ models with c.c. tending to zero to some known, and non-perturbatively well defined, string theory construction in asymptotically flat space.  The problem is that we don't understand any models of quantum gravity in asymptotically flat three dimensional space, even at the level of string perturbation theory.  

However, the {\it a priori} gravitational bounds on the amount of field theoretic entropy that can be accommodated in dS space follow the same dimension dependent scscaling law $ S \sim (R/L_P)^{2/3}$ as in higher dimensions.  Thus we can imagine a model of finite mass fermions and scalars, with attractive interactions, which could make stable complex detectors much smaller than the dS horizon scale.  The center of mass of such a detector would be a semi-classical variable so the spreading of its wave function is irrelevant.  It cannot however remain at the origin of a static patch because Gibbons-Hawking radiation of the light particles, with typical momentum $\sqrt{m/R}$ will cause it to random walk to within a Planck scale of the horizon in a time that is power law in $R$.

There have been attempts to model certain aspects of $dS_3$ by the double scaled SYK model\cite{hv}\cite{suss}.  Since this model has a time independent Hamiltonian, one has to ask what the physical model of the clock could be.  Our discussion above shows that one cannot construct a physical system in semi-classical $dS_3$ that can measure static time for arbitrarily long intervals.  So such Hamiltonians cannot pretend to be any sort of fundamental model of the space-time.  In addition, the model of\cite{hv} allows the deficit angle of localized objects to take values greater than $2\pi$, so it cannot apply to $dS_3$ with spherical spatial slices.  

It's important to point out both the generality of these conclusions (they follow for any relativistic free field equation) and the difference between this scrambling behavior and the scrambling behavior on a black hole horizon in asymptotically flat or AdS space.  In the latter cases, the detector can be localized on the boundary at infinity and can have a clock of limitless precision without violating any of the constraints on the theory.  In dS space, the detector always has to be located inside the static patch.  In $3$ dimensions, with only gravitational interactions, it cannot be a complex object and cannot really serve as a detector, but even its center of mass position cannot be localized in the static patch for more than a limited time. 
\section{$d \geq 4$}

For $d \geq 4$ we have, at least in perturbation theory, a large number of consistent models of quantum gravity in asymptotically flat space.  It has become clear from the AdS/CFT correspondence that the cosmological constant is a model parameter to be chosen, rather than a derived ``ground state energy density" to be calculated, so we might imagine getting a handle on models of dS space by finding sequences of models that converged to some particular flat space model.  

Two obstacles stand in the way of such a program.  The first is that we do not actually know, even in principle, how to construct the non-perturbative Hilbert space of quantum gravity in asymptotically flat space.  Perturbative string theory pretends that it is the Fock space of supergravitons and other stable particles.  We {\it know} that this is wrong when $ d = 4$.  Furthermore, approaching vanishing c.c. from the limit of either positive or negative non-vanishing c.c., one concludes that the ``empty Penrose diagram" state of Minkowski space {\it in any space-time dimension} has infinite entropy. The matrix theory construction\cite{bfss} of eleven dimensional supergravity amplitudes also makes it clear that the question of whether the S matrix is unitary in Fock space is highly non-trivial.  
Understanding the relation of particle physics in dS space to physics on the horizon is obviously crucial, so it seems we will need to understand the non-perturbative theory of flat space better than we do in order to get a handle on dS space.

The second issue has to do with the empirical fact that all known consistent models of asymptotically flat space are exactly supersymmetric, and that supergravity does not admit dS solutions above $d = 4$.  We can draw one of two conclusions from this.  Either there are no consistent models of quantum gravity in $dS_d$ for $d > 4$, or the strategy we proposed at the beginning of this section fails above $d = 4$.

The four dimensional case is of obvious interest for the real world.   Here, two comments are in order.  The first is that, assuming our own universe is evolving towards an asymptotically dS future, it has taught us about the existence of long lived localized excitations, and we have learned, to a large extent, how to model them.  These are called {\it local groups of galaxies}.  The trajectory of the center of mass of such a local group doesn't stay inside the static patch of any particular asymptotic dS geodesic for a time longer than a power of $R_{dS}$, because of Gibbons-Hawking radiation. The clock provided by the motion of that center of mass is decohered by the many constituent particles of the group.  It provides the home for many potential semi-classical q-bits that can measure and store quantum information about the space-time.  The potential lifetime of these detector systems is the time that it takes the group of galaxies to collapse into a black hole.  After that time, the localized q-bits still exist, but information is scrambled between them too rapidly for subsystems to serve as detectors.  Note that the three dimensional detectors discussed above can have a longer useful lifetime because there are no black holes in  $dS_3$.

From a purely mathematical point of view, we can try to find a more precise model of the dS space to which our universe appears to asymptote by matching properties of the standard model to a non-perturbative definition of a string theory model in asymptotically flat space-time.  It is at least plausible that the dimension $4$ and $5$ operators that are responsible for the spectrum of quark and lepton masses and the CKM matrix, can be calculated in such a model, if\cite{oldideas} our model of dS space is part of a sequence that converges to the scattering operator of the asymptotically flat theory.  To get the dS physics right, we would probably have to understand the correct treatment of the soft supergraviton sector of the Minkowski model.  Other aspects of real world particle physics that are likely to be calculable in such a flat space model are the details of baryon decay operators and contributions to lepton number and lepton flavor violating decays.  Other questions, like the resolution of the strong CP problem and the origin of baryogenesis, may or may not have a flat space answer\cite{oldideas}\cite{tbwfbaryo}. 
The main point for the present paper is that the flat space model is, in principle, subject to precise mathematical definition, even if we do not yet know how to do it.  This will probably give us more precise control over models for finite c.c. (of order $10^{-123}$ in Planck units), which converge to it, than a detector in our universe can possibly measure.  Such a detector can in principle have many fewer than $\sim 10^{90}$ semi-classical q-bits, and their lifetime is bounded by that of the longest lived cluster of galaxies.  Furthermore, the detector information is idiosyncratic, and is affected by its formation history and the perturbations exerted on it by other localized excitations, as well as by Gibbons-Hawking radiation.   On the other hand, constraints on the model of dS space coming from matching to a well defined model of asymptotically flat space, are independent of all of the history of our particular universe or any particular detector in it.

This strategy depends on the point of view developed in\cite{oldideas}. Certain aspects of particle physics, mostly associated with the breaking of SUSY, are intimately tied to the existence of a cosmological horizon.  Others are primarily determined by a limiting theoretical model that takes the c.c. to zero and ignores cosmology.  That model is mathematically precise and does not depend on the vagaries of our own cosmological history.  It could be the non-perturbative completion of a known string perturbation series, or a model of M-theory on a $G2$ manifold.  Matching our model of dS space to such a model could give us more information than any experiment we can actually perform in the universe available to us.   

Many modern high energy theorists, under the influence of string theory, have become as interested in purely mathematical issues as the traditional pursuit of matching models to experiment.  There is nothing wrong with that.  It is just valuable to distinguish the two kinds of intellectual pursuit.  String theory models in asymptotically flat and AdS space are (at least in principle in the flat case), mathematically well defined.  The burden of this essay is that the theory of dS space cannot be defined in the same way, if we follow the hints of semi-classical physics and try to model the results of an idealized set of ``experiments".  If we accept the idea of dS space as a finite dimensional quantum system, then all conceivable detectors have a number of semi-classical q-bits that is a power less than one of the total number of q-bits in the space-time, and a lifetime that is bounded by a power less than one of the total number of q-bits.  No quantum theory with a time independent Hamiltonian or an S matrix can describe such a system.  We argued that the most precise mathematical description of the ultimate dS state that our universe appears to be approaching, might be an asymptotically flat superstring model matched to certain features of the standard model.  No set of experiments in our universe could, even in principle, verify every bit of quantum information in such a model, but of course it might lend itself to many experimental checks.

\end{document}